\documentclass[12pt]{article}
\usepackage[whole]{bxcjkjatype} 
\usepackage{multirow}
\usepackage{amssymb}
\usepackage{float}
\usepackage[ruled,vlined]{algorithm2e}
\usepackage{amsmath}
\usepackage{comment}
\usepackage{stmaryrd} 
\usepackage{booktabs}
\usepackage{fancyhdr}
\usepackage{dsfont}
\usepackage{bm}
\usepackage{pifont}
\usepackage{mathtools}
\usepackage[style=numeric]{biblatex}
\usepackage{tikz-cd}
\usepackage{enumitem}
\usepackage{tabu}
\usepackage{geometry}
\usepackage{listings}
\usepackage{color}
\usepackage{authblk}

\lstdefinestyle{mystyle}{
    backgroundcolor=\color{backcolour},   
    commentstyle=\color{codegreen},
    keywordstyle=\color{magenta},
    numberstyle=\tiny\color{codegray},
    stringstyle=\color{codepurple},
    basicstyle=\footnotesize,
    breakatwhitespace=false,         
    breaklines=true,                 
    captionpos=b,                    
    keepspaces=true,                 
    numbers=left,                    
    numbersep=5pt,                  
    showspaces=false,                
    showstringspaces=false,
    showtabs=false,                  
    tabsize=2
}
 
\usepackage{graphicx}
\usepackage{subcaption}
\usepackage[sc]{mathpazo}
\usepackage[T1]{fontenc}

\title{The Evolution of Rumors on a Closed Platform during COVID-19}
\author[1]{Andrea W Wang \thanks{wenyi@iorg.tw}}
\author[2]{Jo-Yu Lan \thanks{m0907618@mail.fcu.edu.tw}}
\author[3]{Chihhao Yu \thanks{chihhao@iorg.tw}}
\author[4]{Ming-Hung Wang \thanks{tonymhwang@gmail.com}}
\affil[1,3]{Information Operations Research Group (IORG)\thanks{https://iorg.tw/}}
\affil[2,4]{Department of Information Engineering and Computer Science, Feng Chia University}

\date{}
\begin{document}

\maketitle
\section{Introduction}

Online social media has democratized contents. By creating a direct path from content producer to consumers, the power of production and sharing of information has been redistributed from limited parties to general populations. However, social media platforms have also given rise to the proliferation of misinformation and enabled the fast dissemination of unverified rumors \cite{vosoughi2018spread} \cite{lazer-2018} \cite{del2016spreading}. In 2020, the COVID-19 pandemic put the world in crisis on both physical and psychological health. Simultaneously, a myriad of unverified information flowed on social media and online outlets. The situation was so severe that the World Health Organization identified it an \textit{infodemic} on February 2020 \cite{who}. According to studies, rumors and claims regarding erroneous health practices can have long-lasting effects on physical and psychological health,  and it even interfered with the control of COVID-19 in various parts of the world \cite{abdoli_2020} \cite{tasnim2020impact}. 

In light of the \textit{infodemic}, several investigations have been carried out to look at the COVID-19 misinformation issue in various aspects. Topics included but not limited to, the types and contents of COVID-19 misinformation \cite{yang2020prevalence} \cite{brennen2020types}, the spread and prevalence of rumors on social media platforms \cite{cinelli2020covid},  \cite{kouzy_coronavirus},  \cite{gallotti2020assessing}, \cite{yang2020prevalence}, \cite{pulido-more-retweets}, \cite{shahi_exploratory_2021}, the consequences of misinformation \cite{bridgman_causes_2020}, and the application of machine learning algorithms on rumor analyses \cite{shi_rumor_2020} \cite{jelodar2020deep}. However, the majority of the studies focused on data collected from public social media platforms such as Twitter, Facebook, or Weibo. Explorations on closed messaging platforms, such as WhatsApp, WeChat, or LINE, remained extremely scarce. While popular social media platforms are indeed important targets to study online behaviours and expressions, closed platforms remain an integral place to look at, given its more private settings.

Our contribution to the current research is in three ways. First, by investigating COVID-19 messages on LINE, we added to the limited research of COVID-19 rumors on closed messaging platforms \cite{ng2020analysing} \cite{p_covid-19_2020}.  According to the survey by Taiwan Communication Survey in 2018, $98.5\%$ of people in Taiwan used LINE as their primary messaging tool, making LINE the most popular instant message platform in Taiwan.\footnote{Data were
collected by the research project of the Taiwan Communication Survey
(TCS), which is supported by the Ministry of Science and Technology
of R.O.C. The author(s) appreciate the assistance in providing data by
the institute aforementioned. The views expressed herein are the
authors' own. doi: 10.6141/TW-SRDA-D00176-1} We looked into a dataset of $114,124$ suspicious messages reported by LINE users in Taiwan between January, 2020 to July, 2020 

Secondly, we proposed an efficient algorithm that could cluster a large number of text messages according their topics and narratives without having to decide how many groups beforehand. The results were clusters where each one only contains messages that are within limited alterations among each other. Thus,  each cluster is one specific rumor. 

Third, by using the results from the algorithm, we were able to look at the dynamics of each particular rumor over time. To the best of our knowledge, we are the first to study not only how the content of a specific COVID-19 rumor evolved over time but the interaction between content change and popularity. We found that some form of content alterations were successful in aiding the spread of false information. 

The major findings of this work are three-fold:

\begin{enumerate}
    \item By combining Hierarchical Clustering and K-Nearest Neighbors, we could reduce computational time of clustering to linear time. This would enable the large-scale study of rumor transformation.
    \item Fact-check did not effectively alleviate the spread of COVID-19-related false information. In fact, the popularity of rumors were more influenced by major societal events. 
    \item Key authoritative figures were often falsely mentioned or quoted in misinformation, and such practice helped with the popularity of a message.
\end{enumerate}

This paper is organized as followed: we introduced our data in Section \ref{sec: data}. Next, we presented the proposed algorithm to cluster text data in Section \ref{sec: method} and subsequently compared the proposed algorithm with other clustering techniques in Section \ref{sec: alg-result}. Finally we reviewed 3 high-volume  COVID-19 false information in Section \ref{sec: cases}. We discussed and concluded this work in Section \ref{sec: discussion} and \ref{sec: conclusion}.

In the following sections, we used \textit{clusters} and \textit{groups} interchangeably. And we described a group of suspicious messages as one \textit{rumor}, since belonging to the same group meaning they were seen as one narrative. And then we referred to rumors that are verified false as \textit{misinformation} or \textit{false information}.

\section{Related Works}
From the inception of the pandemic, several survey studies revealed that people relied on social media to gather COVID-19 information and guidelines \cite{mat2021attitude} \cite{mubeen2020knowledge}. Misinformation on social media has since been a keen interest of the research community. 

Efforts have been put into studies of true and false rumors on social media \cite{pulido-more-retweets}.  For example, Cinelli et al. compared feedbacks to the reliable and questionable information across five platforms, including Twitter, YouTube, and Gab. The study showed that users on the less regulated platform, Gab, responded to questionable information 4 times more than those on the reliable ones.  YouTube users were more attracted to reliable contents, and Twitter users reacted to both contents more equally  \cite{cinelli2020covid}.  Gallotti et al. looked at the how much unreliable information Twitter users were exposed to across countries. While the level of exposure was country dependant, they revealed that the exposure to unreliable information decreased globally as the pandemic aggravated \cite{gallotti2020assessing}. 

Machine learning and deep learning techniques have been used to study the topics and sentiments for COVID-19 misinformation \cite{abd2020top}. For example, Jelodar et al. used Latent Dirichlet Allocation to extract topics from $560$ thousands of COVID-19 Twitter posts and then used LSTM neural network to classify sentiments of posts \cite{jelodar2020deep}. By applying Structure Topic Model and Walktrap Algorithm, Jo et al. classified questions and answers from South Korea's largest online forum and discovered that questions related to COVID-19 symptoms and related government policies revealed the most fear and anxiety \cite{jo2020online}. Furthermore, by employing a multimodal deep neural network for demographic inference and VADER model for sentiment analysis, Zhang et al. performed a cross sectional study on Twitter users. They found that older people exhibited more fear and depression toward COVID-19 than their younger counterparts, and females were generally less concerned about the pandemic \cite{zhang2021understanding}. 

Previous investigations on rumors indicated that individuals are more likely to believe in questionable statements after seeing repeatedly \cite{boehm1994validity} \cite{berinsky2017rumors}, and that rumors became more powerful after being shared multiple times \cite{difonzo2007rumor}. Most studies only look at the broad topics of misinformation. For example, some looked at reliable versus unreliable information \cite{cinelli2020covid} \cite{gallotti2020assessing} \cite{yang2020prevalence}, and others employed natural language processing techniques to reduce thousands of social media posts into $10$ to $20$ groups of topics \cite{abd2020top} \cite{jelodar2020deep} \cite{jo2020online} \cite{cinelli2020covid}. Shih et al. instead investigated the content change and temporal diffusion pattern of $17$ popular political rumors on twitter \cite{shin2018diffusion}. They found that false rumors came back repeatedly, usually becoming more extreme and intense in wordings, while true information did not resurface at all.  To the best of our knowledge, there has not been similar study at COVID-19 rumors.

\section{Data} \label{sec: data}

In Taiwan, LINE users can voluntarily forward suspicious messages to fact-checking LINE bots such as Cofacts \footnote{https://cofacts.tw/} or MyGoPen \footnote{https://mygopen.com/}. The bots archive the messages and check against their existing databases. If such message has been fact-checked, the bots would reply with the fact-checked results.

We obtained a dataset of $210,221$ suspicious messages forwarded by LINE users to a fact-checking LINE bot between January to July, 2020. The dataset included rumors related to COVID-19 and also some other topics. To do clustering, we preprocessed each message by the following steps: 
\begin{enumerate}
    \item Removed non-Simplified or non-Traditional Chinese Characters.
    \item Tokenized with Jieba \footnote{https://github.com/fxsjy/jieba}.
    \item Removed tokens that are Chinese stopwords.
\end{enumerate}

In the following sections, we focused on longer texts. We only looked at $114,124$ messages having at least $20$ tokens. The character distributions is presented in Table \ref{tab: data-char-long}.

Along with the text content of each reported message, we also obtained the report time of each message and a unique identifier for the LINE user that reported the message. It is to note that the user identifier we received were scrambled, therefore, it was not possible for us to use the identifiers to attribute any message back to any actual LINE user.

\begin{table}[H]
    \centering
\begin{tabular}{lrrrrrr}
\toprule
       & All & Chinese & Digits & English & Others & Number of \\
       && Characters & & Alphabets & &Tokens \\
\midrule
Min & 24 & 24 & 0 & 0 & 0 & 20 \\
Median & 233 & 145 & 7 & 2 & 38 & 58\\
Max & 10012 & 8132 & 3252 & 7014 & 5532 & 2971 \\

\bottomrule
\end{tabular}
    \caption{Characters components of messages having at least 20 tokens. "Others" include characters such as punctuation marks and emojis.}
    \label{tab: data-char-long}
\end{table}

\begin{figure}[H]
    \centering
    \includegraphics[width =1\textwidth]{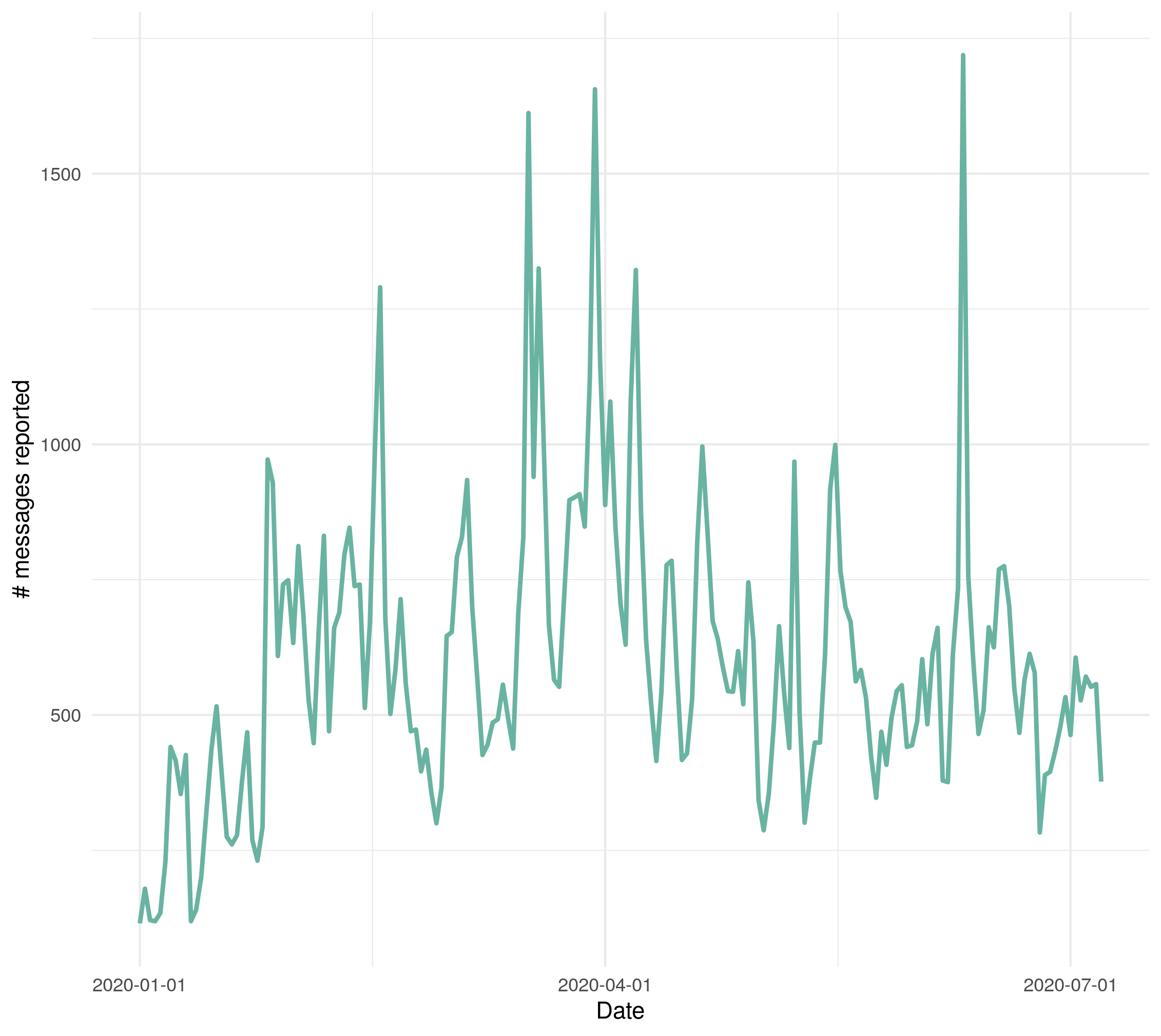}
    \caption{number of suspicious messages reported by date}
    \label{fig:data-size-by-month}
\end{figure}

\section{Method} \label{sec: method}
In this section, we described our problem and the proposed clustering algorithm. To follow the terminology of Natural Language Processing, in this section we used \textit{document} to refer to one \textit{message} in our dataset.

\subsection{Problem Definition}
Given a set of $n$ documents, we would like to group them into $m$ clusters, of which each cluster are made up of documents very similar in usage of terms, only within a limited degrees of text alterations. Intuitively, we wanted the same cluster to have documents that \textit{talked about the same thing in the same way}. Note that $m$ is unknown beforehand. 

For example, given two documents A and B, they should be in the same cluster if the overlapping terms of A and B constitute a large part of both A and B. However, if the overlapping terms make up a large part of A but not B, then they should be in different clusters, because that means B is made up of A and also some other terms.

Formally, we defined the terms in a document to be its token set after tokenization. And the distance between two documents A and B to be 
\begin{equation}\label{formula: dist-metric}
\mathbold{d}(A, B) = 1 - \frac{|tok(A) \cap tok(B)|}{max(|tok(A)|, |tok(B)|)}
\end{equation}
where $tok(\cdot)$ is the set of tokens of one document. And $|\cdot|$ is the number of elements in a set. 

\subsection{The Cluster-Classification, "Hybrid",  Algorithm}

\begin{figure}[H]
    \includegraphics[width =1\textwidth]{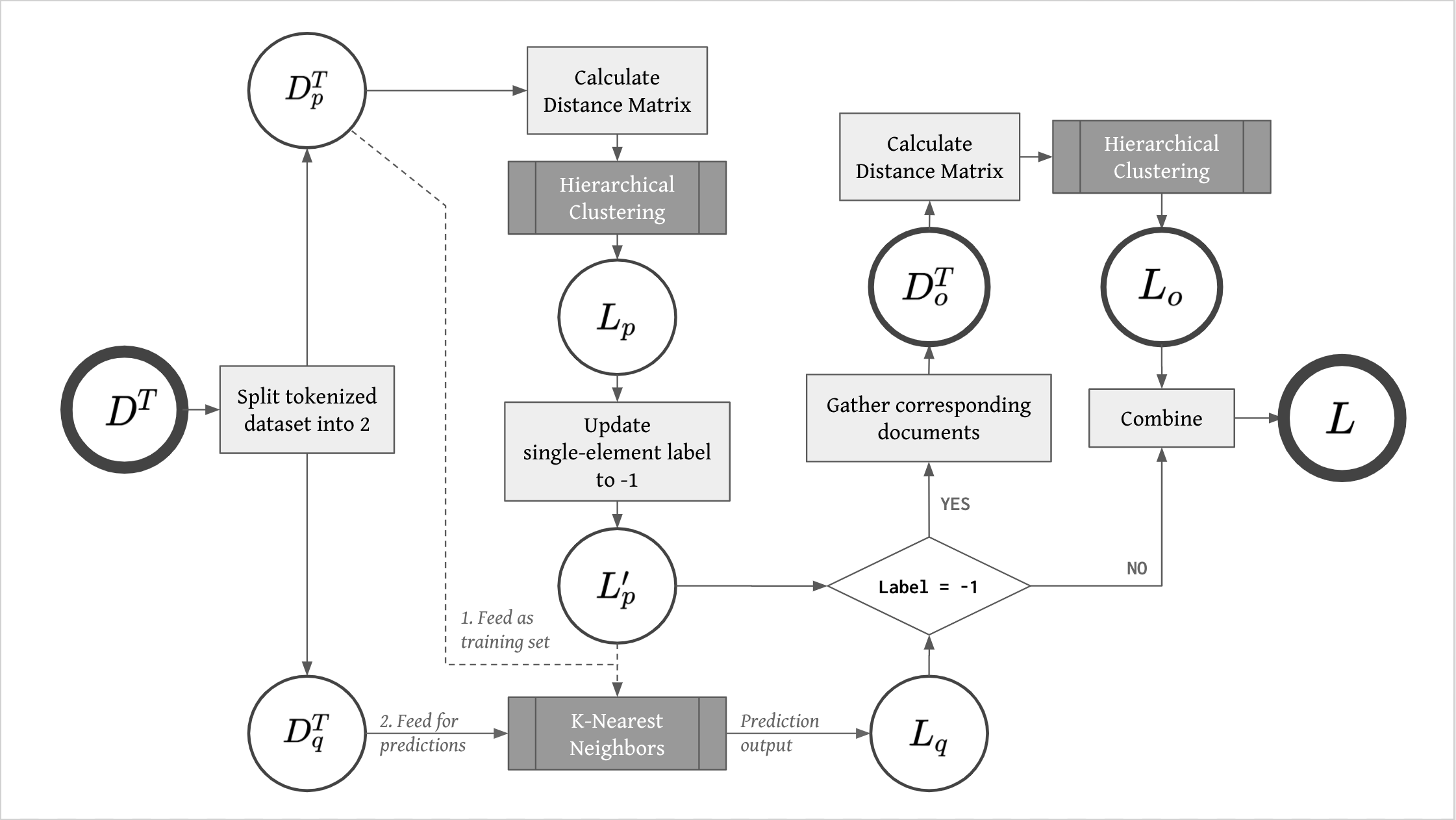}
    \caption{Algorithm flow diagram}
    \label{fig:algo-flow}
\end{figure} 

\paragraph{Notation}
\begin{enumerate}
    \item $(A)_j$: $j^{th}$ element of set $A$.
     \item $Label(x)$: The label of element $x$.
\end{enumerate}

\paragraph{Input}
\begin{enumerate}
\item $D$: the set of all documents to be grouped. 
\item $D^T$: the set of tokenized documents. Each element $(D^T)_i$ is the token set of document $(D)_i$.
\item \textit{train portion} $\mathbold{p}$: a number in $(0, 1]$.
\item \textit{distance threshold} $\mathbold{\lambda}$: a number in $(0, 1]$.
\end{enumerate}

\paragraph{Algorithm} 
\begin{enumerate}
    \item Select $\mathbold{p} \times |D^T|$ elements from $D^T$, denoted as $D^T_p$, and the rest not selected as set $D^T_q$.
    \item \label{alg: make-dist-matrix} Construct distance matrix $M$ for $D^T_p$, where $M_{i,j} = \mathbold{d}((D^T_p)_i, (D^T_p)_j)$ by Formula \ref{formula: dist-metric}. Note that $M$ is symmetric.
    \item \label{alg: clustering} Feed $M$ into Hierarchical Clustering with distance threshold of $\mathbold{\lambda}$. We would get back a sequence of numbers $L_p$, where $(L_p)_i$ is the label of element $(D^T_p)_i$. Elements with the same label are in the same cluster. Since the number itself does not carry meaning, manipulate them so they are all non-negative whole numbers. 
    \item \label{relabel} $\forall (L_p)_i \in L_p$, if $|\{l | l=(L_p)_i \forall l \in L_p\}| = 1$, then replace the value of $(L_p)_i$ to $-1$. Denote the updated label set as $L_p^\prime$.
    \item Train a K-Nearest Neighbors classifier $\mathbold{K}$ using the training set ($D^T_p$, $L_p^\prime$). And then use $\mathbold{K}$ to predict the labels of $D^T_q$. Denote the prediction as $L_{q}$.
\item  \label{alg: construct-label} Construct $L$ from $L_p^\prime$ and $L_{q}$, where $(L)_i = Label((D^T)_i)$.
    \item Construct $D^T_O = \{d_i| Label(d_i) = -1 \forall d_i \in D^T\}$. 
    \item Redo step \ref{alg: make-dist-matrix} and \ref{alg: clustering} for $D^T_o$. Denote the resulting sequence as $L_o$. Make sure the values of $L_o$ do not overlap with the values of $L$ from step \ref{alg: construct-label}.
    \item Update $L$ from step \ref{alg: construct-label} with $L_o$. 
\end{enumerate}

\paragraph{Output}
Output is $L$. The $i^{th}$ element of $L$, denoted as $(L)_i$, is the label of $(D^T)_i$. Note that the value of the label itself does not carry any meaning. However, elements in $D^T$ with the same label belong to the same cluster.

\section{Results}

\subsection{Ground truth}
We randomly selected $50,000$ messages from the dataset and used pure Hierarchical Clustering algorithm to perform clustering. The messages were separated into $7,401$ groups. The largest group had $1,082$ messages, and the smallest group contained only $1$. There were $5,231$ groups with only 1 message, meaning the rest of $44,796$ messages were separated into $2170$ groups. There were $12$ groups with at least $500$ messages.

\begin{table}[H]
    \centering
\begin{tabular}{p{0.3\linewidth}rrrrrr}
\toprule
       & mean & std & max & $Q_3$ & $Q_2$ & min \\
\midrule
All Groups & 6.756 & 39.190 & 1082 & 2 & 1 & 1\\
Groups with at least 2 elements & 20.631 & 70.478 & 1082 & 10 & 3 & 2 \\
\bottomrule
\end{tabular}
    \caption{Group size statistics}
    \label{tab: result-ground-truth-group-size-stat}
\end{table}

\subsection{Model Comparisons} \label{sec: alg-result}
\subsubsection{Evaluation Metrics}
We opted precision, recall and F-score as evaluation metrics. In the sense of information retrieval, precision is the number of correct results returned divided by all results returned from search. Hence, high precision means the predictions are very relevant. On the other hand, \textit{recall} measures the number of correct results returned divided by the total number of correct results.  High recall corresponds to the completeness of returned results. Note that simply by returning all documents, one could achieve $100\%$ of recall, but that will result in very low precision. Therefore, precision and recall need to be taken together to determine the quality of classification. F-score, defined as the harmonic mean of precision and recall, is one such measure that combine precision and recall. 

\subsubsection{Experiments Settings}
We compared speed and performances among 4 models:
\begin{enumerate}
    \item Hierarchical Clustering only (\textbf{clustering}). The result from this model is considered to be ground truth. 
    \item Cluster-Classification Model (\textbf{hybrid}). This is our proposed algorithm.
    \item Latent Dirichlet Allocation (\textbf{LDA}).
    \item KMeans with PCA dimensionality reduction (\textbf{pca+kmeans}).
\end{enumerate}
Throughout the experiments we used distance threshold $\mathbold{\lambda} = 0.6$.

Both \textbf{LDA} and \textbf{pca+kmeans} clustering required a predefined number of groups, which doesn't really fit out purposes. However, for the sake of comparison, we would use the number of groups outputted by \textbf{clustering} model as input to both models.

\subsubsection{Measuring model performances}
Suppose the input is tokenized set of $k$ documents $D^T$ and the \textbf{clustering} model put $k$ documents into $n$ groups, ($g_1$, $g_2$, ... $g_n$). $g_1$ is the group having largest number of documents and $g_n$ the least.  Another model $M$ put $D^T$ into $m$ groups: ($l_1$, $l_2$, ..., $l_m$). We calculated precision, recall and F-score of model $M$ by the following algorithm:

\begin{algorithm}[H]
 $i\gets 1$, $c \gets 0$, $p \gets 0$, $r \gets 0$, $f \gets 0$\;
 \While{$c < k/2$}{
  Find $l_k$ where $l_k$ has the most overlapping components with $g_i$\; 
  calculate precision $p_k$, recall $r_k$, and F-score $f_k$ of $l_k$ by comparing with $g_i$\;
  $r \gets r + r_k$\;
  $p \gets p + p_k$\;
  $f \gets f + f_k$\;
  $i \gets i + 1$\;
  $c \gets c + |g_i|$ \;
 }
 \KwResult{
    precision $\gets p/i$\;
    recall $\gets r/i$ \;
    F-score $\gets f/i$ \;
 }
 \caption{Calculating Precision, Recall, F-score}
\end{algorithm}

In each experiments, we did $5$ iterations. In each iteration, we randomly selected $k$ messages from our dataset. We would get $1$ precision and recall after each iteration, and we used the results of $5$ iterations to calculate confidence intervals. 

\subsubsection{Experiments Results}
\begin{figure}[H]
    \centering
    \includegraphics[width = 1\textwidth]{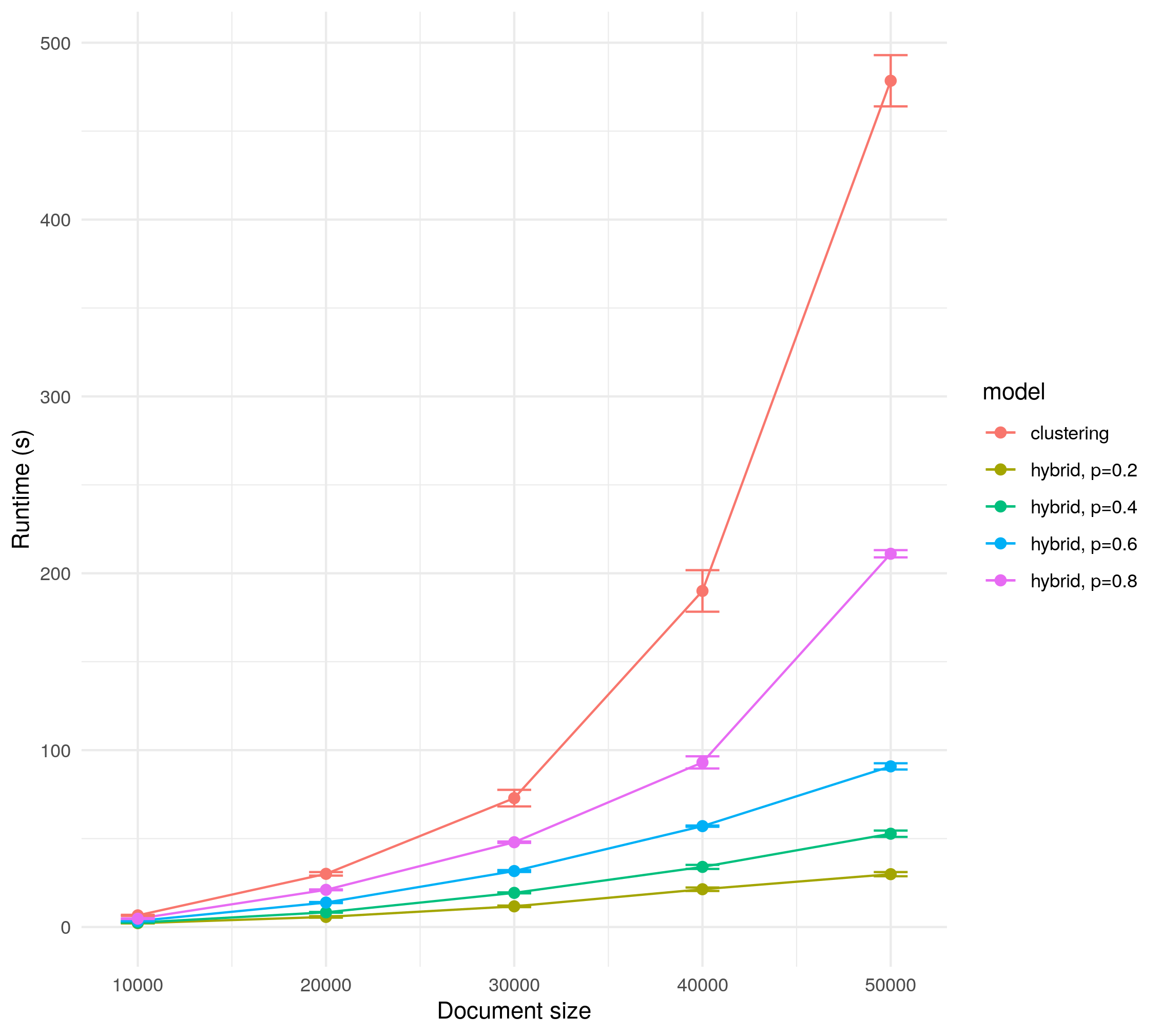}
    \caption{speed comparison between \textbf{clustering} and \textbf{hybrid} across different levels of $\mathbold{p}$. Using \textbf{hybrid} with $\mathbold{p}$ lower than $0.6$ reduced the runtime from exponential to linear time. }
    \label{fig:cluster-cc-p-time}
\end{figure}

\begin{figure}[H]
    \centering
    \includegraphics[width = 1\textwidth]{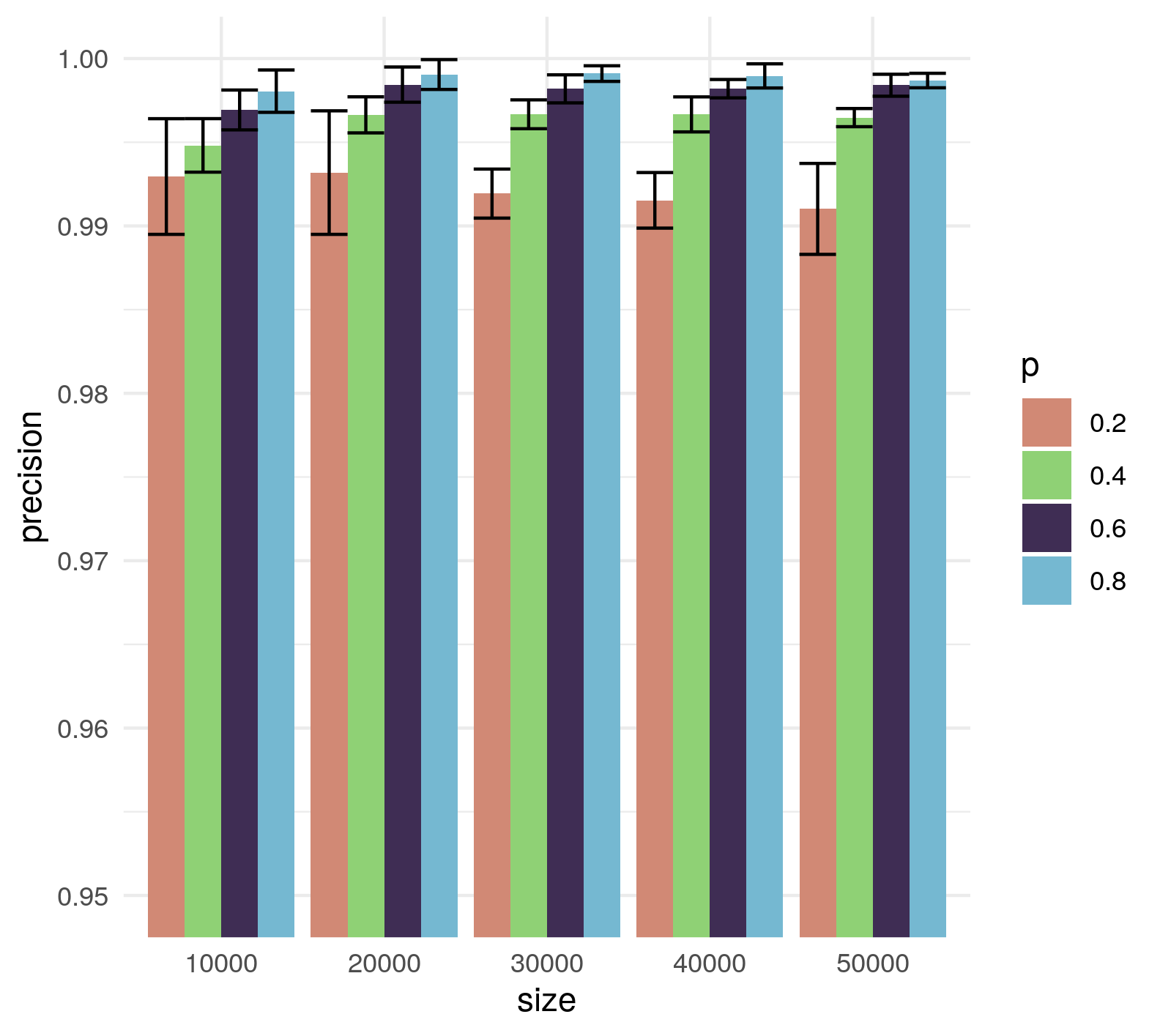}
    \caption{Precision}
    \label{fig:precision-p}
\end{figure}

\begin{figure}[H]
    \centering
    \includegraphics[width = 1\textwidth]{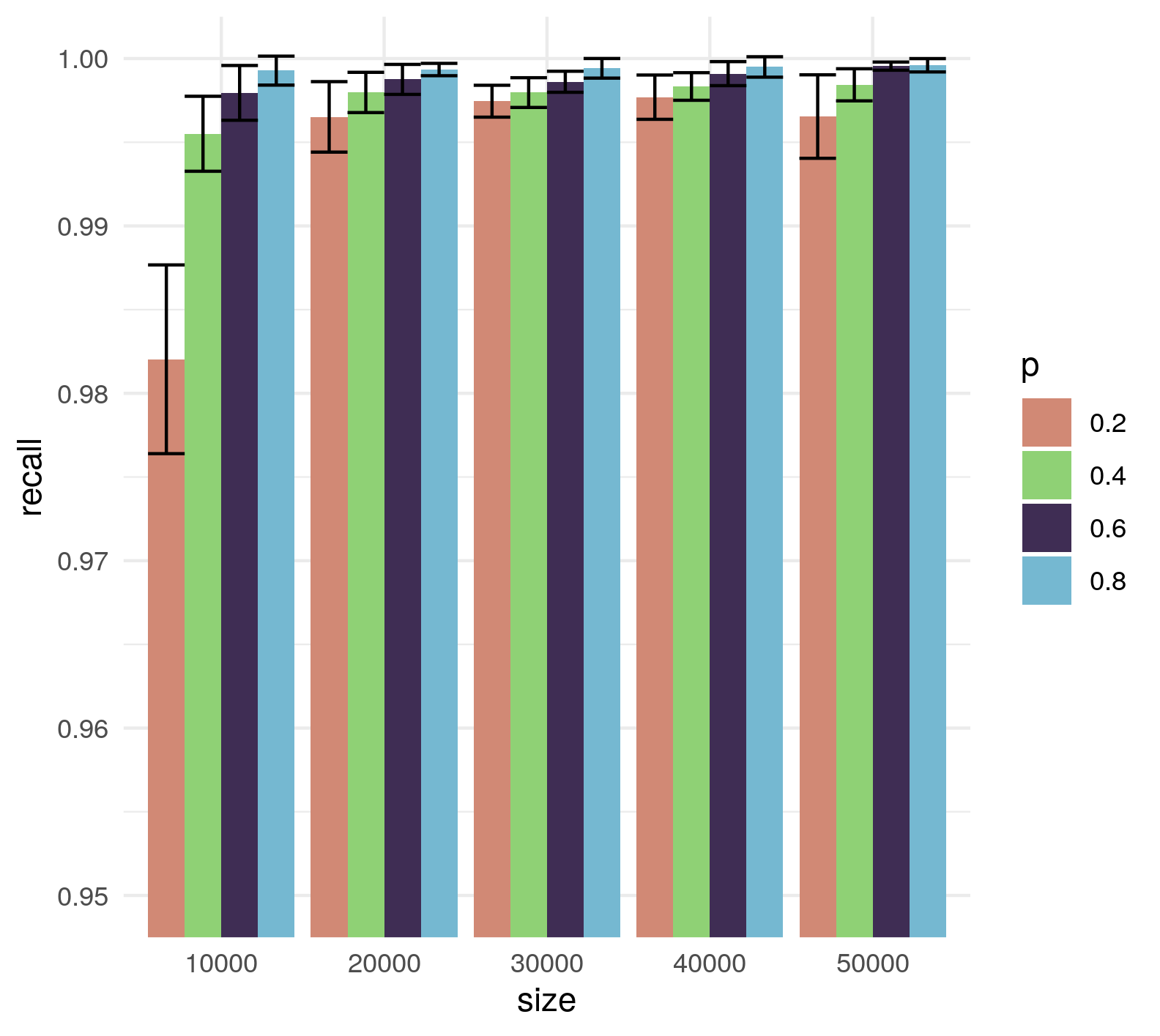}
    \caption{Recall}
    \label{fig:recall-p}
\end{figure}

\begin{figure}[H]
    \centering
     \includegraphics[width = 1\textwidth]{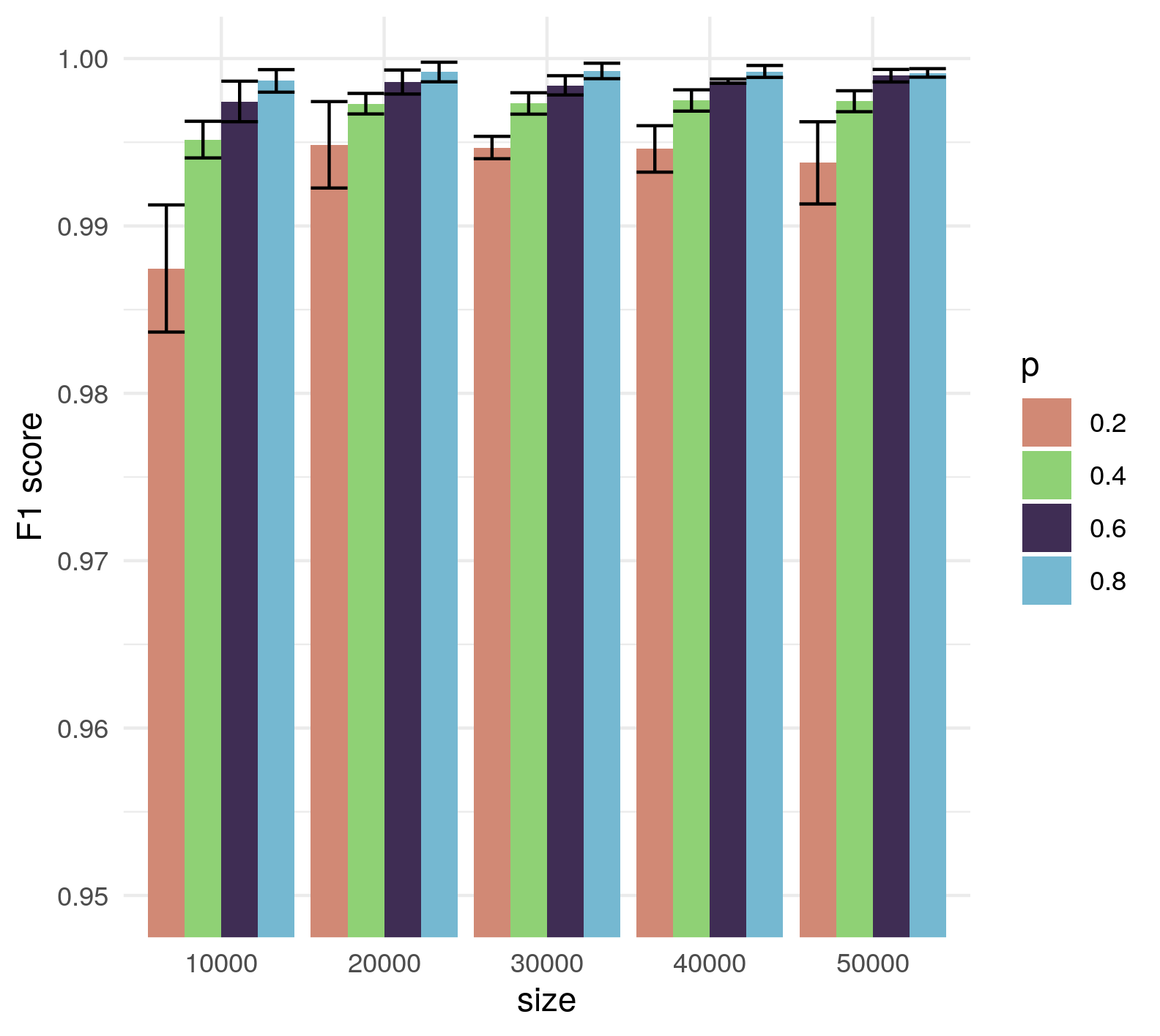}
    \caption{F-Score}
    \label{fig:f-score-p}
\end{figure}

As shown in Figure \ref{fig:cluster-cc-p-time}, the \textbf{hybrid} model greatly reduced the time required especially when $\mathbold{p}$ was equal or less than $0.6$. Furthermore, the performance metrics remained greater than $99\%$ across levels of $\mathbold{p}$ (Figures \ref{fig:precision-p}, \ref{fig:recall-p}, \ref{fig:f-score-p}). It showed that the \textbf{hybrid} model's assignments of groups were very complete (measured by recall), and that the classification of K-Nearest Neighbors did not introduce too much errors in each group (measured by precision). From Table \ref{tab:perf}, we observed that \textbf{LDA} is much slower that other models. Furthermore, the precision was very low, meaning that predicted groups could have many false positives.  On the other hand, \textbf{pca+kmeans} were 10 times slower than \textbf{clustering}. While the precision was comparable to that of \textbf{hybrid} methods, recall was only $73\%$. This showed that \textbf{pca+kmeans} would miss out many transformations of a message.

\begin{table}[H]
    \centering
\begin{tabular}{lrrrr}
\toprule
   Model & Runtime (s) & Precision  &Recall& F-score \\
         &     mean &      mean &    mean &    mean  \\
\midrule
 clustering &    6.594 &   - &    - &  -  \\
hybrid, $\mathbold{p}=0.2$ &    2.172  &     0.993&  0.982 &   0.986 \\
hybrid, $\mathbold{p}=0.4$ &    2.502 &     0.995 &  0.996&   0.995\\
hybrid, $\mathbold{p}=0.6$ &    3.418 &     0.997 &  0.998  &   0.997 \\
hybrid, $\mathbold{p}=0.8$ &    4.697  &     0.998  &  0.999 &   0.999  \\
LDA & 1788.981  &     0.624 &  0.939 &   0.704 \\
pca+kmeans &   41.143 &     0.993 &  0.734  &   0.823 \\
\bottomrule
\end{tabular}
    \caption{Performance comparison (10,000 documents)}
    \label{tab:perf}
\end{table}

\subsubsection{Clustering 114K messages using the \textbf{hybrid} method}\label{all-grouping-report}
We used hybrid methods with train portion $\mathbold{p} = 0.4$ and distance threshold $\mathbold{\lambda} = 0.6$ to cluster the whole set of $114$ thousands messages. The messages were separated into $12,260$ groups. Among those, $8,529$ groups only had $1$ message. Therefore, the rest of $105,595$ messages were separated into $3,731$ groups. The largest group had $2,546$ messages. There were $15$ groups with at least $1000$ elements. We presented the statistics of group sizes in Table \ref{tab: result-114K-group-size-stat}

\begin{table}[H]
    \centering
\begin{tabular}{p{0.3\linewidth}rrrrrr}
\toprule
       & mean & std & max & $Q_3$ & $Q_2$ & min \\
\midrule
All & 9.309 & 71 & 2546 & 2 & 1 & 1\\
Groups with at least 2 elements & 28.302 & 126.907 & 2546 & 10 & 3&2\\ 
\bottomrule
\end{tabular}
    \caption{Group Size statistics}
    \label{tab: result-114K-group-size-stat}
\end{table}

\subsection{Case Studies} \label{sec: cases}
In this section we presented some high-volume suspicious messages related to COVID-19, obtained from the previous section \ref{all-grouping-report}. 
\subsubsection{Case 1: Do not go outside!}
\begin{table}[H]
    \centering
\begin{tabular}{p{0.4\linewidth} p{0.4\linewidth}}
\toprule
       English Translation & Original\\
\midrule
Academian Zhong, Nan-Shan emphasized repeatedly, 'Do not go outside! Wait until at least the Lantern Festival to assess the situation of the epidemic.' Be warned that even if you're cured, you would suffer the rest of your life. This is a plague worse than SARS. The side effect of the drugs are more severe...This is a war, not a game ... There is no outsider in this war ... & 鐘南山院士再⁠次強⁠調：別出門，元宵後，再看疫情控制情況！警告：一旦染上，⁠就算治癒⁠了，後遺症也會拖⁠累後半⁠生！這場瘟⁠疫比17年⁠前的非典更嚴重，用的藥⁠副作用更大。如果出了特效藥⁠，也⁠只能保命，僅此⁠⁠而已！出⁠⁠門前想想你的家人，別連累⁠家人，⁠⁠⁠能⁠不出門就不出⁠門，大家一起⁠轉發⁠吧！這是一場⁠戰役⁠，不⁠是兒⁠戲，⁠收⁠起⁠⁠你盲⁠目的自信⁠和僥倖⁠⁠心⁠⁠理⁠，⁠⁠也⁠收起你事不⁠關己高高掛起的態度⁠，在這場戰役中沒有⁠⁠局外人！
在⁠家！⁠在家！在家！⁠不要點贊！⁠求轉發  ⁠ ——  鐘南山
\\
\bottomrule
\end{tabular}
    \caption{Case 1 Message Content}
    \label{tab:c1-content}
\end{table}

\begin{figure}[H]
    \centering
    \includegraphics[width = 1\textwidth]{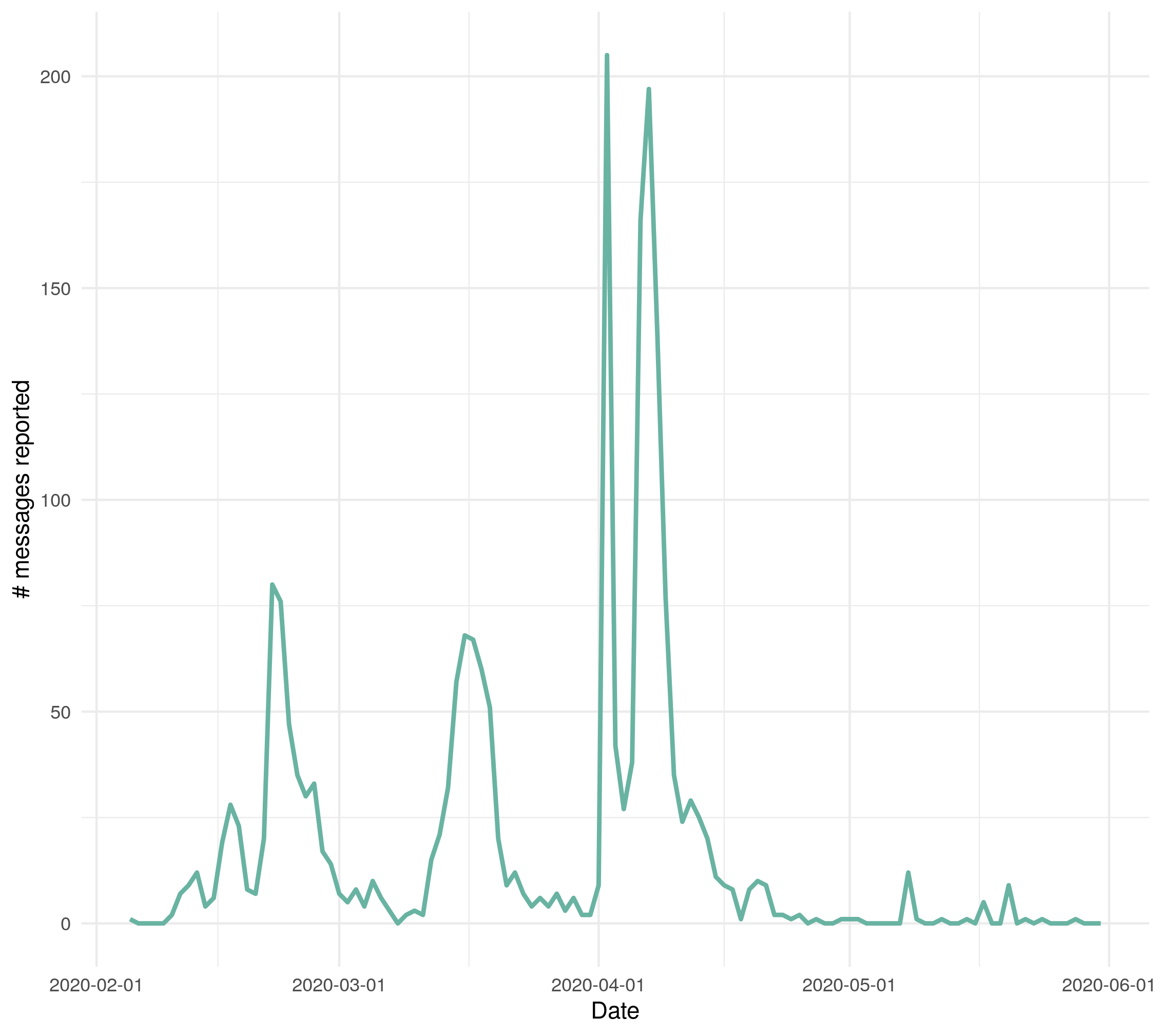}
    \caption{number of documents of Case 1 reported by date. The number peaked on Apr $2^{nd}$ ($205$ documents), the day the Ministry of Health and Welfare announced that this was a misinformation. We subsequently saw another peak on Apr $6^{th}$ (166 documents), the day after a 4-day long weekend.}
    \label{fig:c1}
\end{figure}
This case first appeared in the dataset on Feb $2^{nd}$, 2020. Over the course of 3 and a half months, there were a total of $2,119$ messages reported. The reporting went viral at least four times: it peaked on February $22^{nd}$ ($80$ documents), March $16^{th}$ (68 documents), welcomed the highest peak on Apr $2^{nd}$ ($205$ documents), then the last one on Apr 6th with 166 documents. We observed a number of key characteristic changes in the texts itself over the life of this message. 

First of all, the time-sensitive information in the message evolved with time. At its early stage, "Lantern Festival", on Feb $8^{th}$ in 2020, was spotted in the majority of messages. However, on Feb $18^{th}$, we spotted the first message that replaced "Lantern Festival" with "March". Then, after March $10^{th}$, the majority of reported messages used "Mid-Autumn Festival (June $25^{th}$, 2020)".

Secondly, the efforts were put to emphasize the authoritativeness from whom the message was \textit{quoted}. The first form of this message started with quotation from \textit{The Main-land Academian Zhong, Nan-Shan}, who gained fame during the SARS pandemic in 2003 \footnote{https://en.wikipedia.org/wiki/Zhong\_Nanshan}. Other titles, such as "Expert in Pandemic from Mainland China" or "Expert in Coronavirus", were also observed in some transformations. Then later, on Feb $18^{th}$, age was first seen in the message: \textit{"Expert in Coronavirus from Mainland China, 78-year-old Academian Zhong, Nan-Shan, emphasized..."}. Starting March $10^{th}$ to March $31^{st}$, almost every message included age.  Then starting from April $1^{st}$, every reported message has Zhong replaced by Chen, Shih-chung. As the Director of Taiwan's Central Epidemic Command Center (CECC), Chen's popularity has skyrocketed during the pandemic through his daily press conference. This was also when we observed the highest peaks of the reported messages.

Due to the prevalence of this message spreading on web and closed platforms, the Ministry of Health and Welfare as well as CECC sent out a press release and a facebook post \footnote{https://www.mohw.gov.tw/cp-4633-52577-1.html} \footnote{https://www.facebook.com/470265436473213/posts/1524703107696102/} on April $2^{nd}$, reminding the public that this was a false information. Nevertheless, this did not stop another viral spread of the same message at the end of a four-day long holiday in Taiwan, where crowds were seen in every tourists attraction on the island. For days people were worried that the long-weekend would lead to another outbreak of the pandemic, which explained why the message bearing the key topic "do not go out" would become a big hit. 

\begin{table}[H]
    \centering
\begin{tabular}{{lp{0.4\linewidth} p{0.4\linewidth}}}
\toprule
       Date &  Previous & New\\
\midrule
Feb 17, 2020 & Academian Zhong, Nan-Shan stressed again

鍾南山院士再次強調 & 

\textbf{Pandemic expert from Mainland China}, Academian Zhong, Nan-Shan stressed again

大陸防疫專家鍾南山院士再次強調 \\
\cmidrule{3-3}
Feb 18, 2020 &&
\textbf{Coronavirus expert from Mainland China, 78-year-old} Academian Zhong, Nan-Shan stressed again

大陸，冠狀病毒專家
鐘南山78歲院士再次強調 \\
\cmidrule{3-3}
Feb 27, 2020 && 

Coronavirus expert from Mainland China, \textbf{84-year-old} Academian Zhong, Nan-Shan stressed again

大陸，冠狀病毒專家鐘南山84歲院士再次強調\\
\cmidrule{3-3}

Apr 1st, 2020& & 
Director of Taiwan's Ministry of Health and Welfare, Chen, Shih-Chung, reminded everyone

台灣 衛福部長 陳時中提醒大家\\
\midrule
Feb 18, 2020 & 
Do not go outside! Wait until \textbf{the Lantern Festival} to reassess pandemic situation. 

別出門，元宵後，再看疫情控制情況 & 
Do not go outside! Wait until \textbf{March} to reassess pandemic situation. 
別出門，三月後 再看疫情控制情況\\
\cmidrule{3-3}
&& 
Do not go outside! Wait until the \textbf{Mid-Autumn Festival} to reassess pandemic situation. 

別出門，端午節過後，再看疫情﻿控制情況\\

\bottomrule
\end{tabular}
    \caption{Content Change Log for Case 1}
    \label{tab:c1-changelog}
\end{table}

\subsubsection{Case 2: Drink salty water can prevent the spread of COVID-19.}
In this case we looked at the messages that promoted drinking salt water to prevent the coronavirus. In fact, we investigated two messages and the combination of the them (Table \ref{tab:c2-content}). 

\begin{figure}[H]
    \centering
    \includegraphics[width = 1\textwidth]{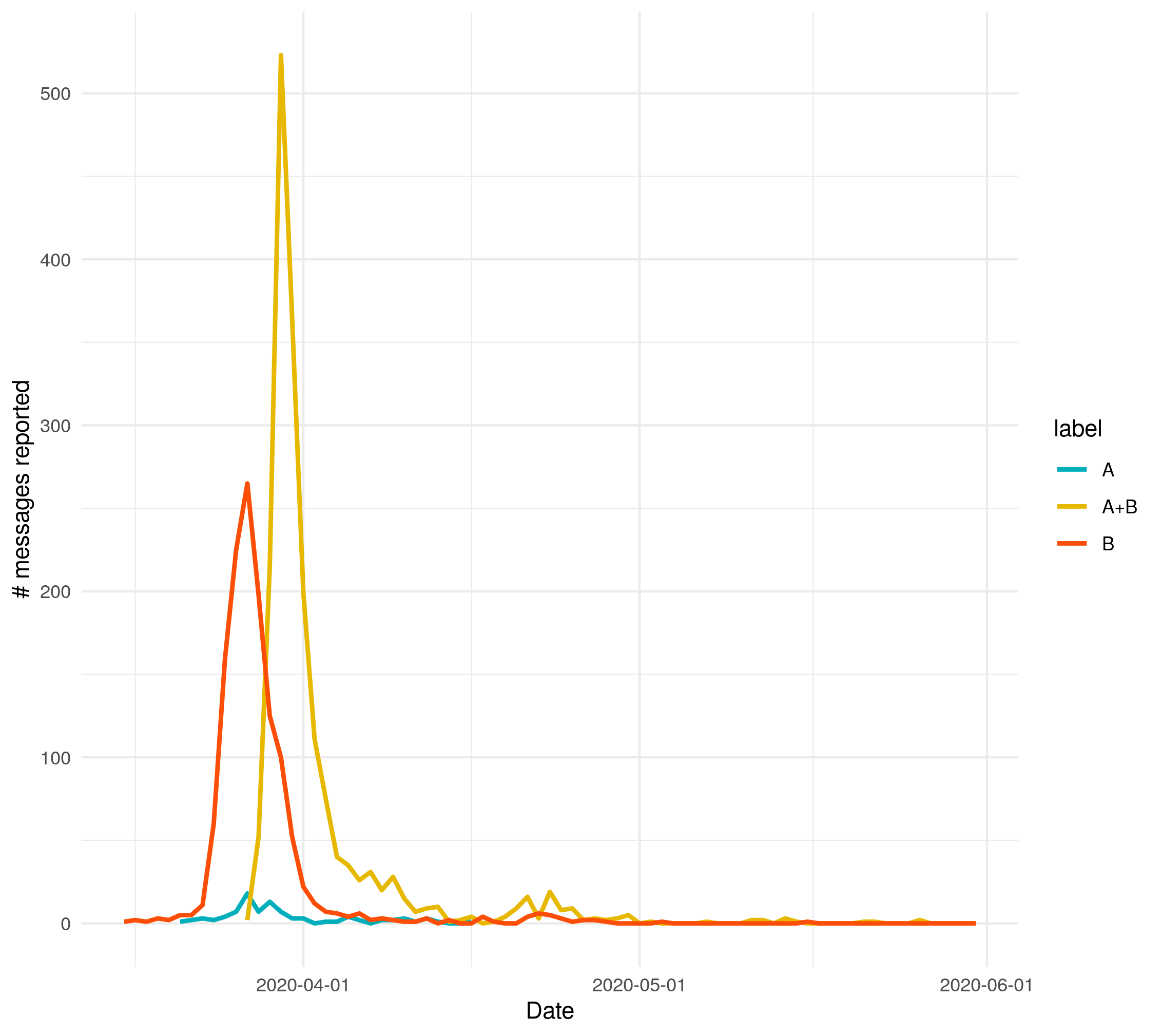}
    \caption{number of documents of Case 2 reported by date. }
    \label{fig:c2}
\end{figure}

We first observed Message (B) in our dataset on March $16^{th}$. Over the course of its evolution, several medical personnel, such as \textit{Director of The Veteran Hospital} or \textit{Dr. Wang of Tung Hospital} (who, in fact, is an Orthopedist), were misquoted. This showed the use of authoritative power to spread this piece of false medical information. The highest peak was on March $27^{th}$, where 265 documents were reported. Around the same time, a small number of Message (A) were also lurking, however, it did not get as much attention as Message (B) before both messages merged into 1 on March $27^{th}$ and went viral shortly after on March $30^{th}$ (Orange line in Figure \ref{fig:c2}). 
In fact, Message (B) was fact-checked by Taiwan FactChecking Center \footnote{https://tfc-taiwan.org.tw/} rather early, on March $19^{th}$ \footnote{https://tfc-taiwan.org.tw/articles/3207} and announced it a misinformation, however, this did not stop the piece from misquoting doctors and continued spreading. As a matter of fact, several translations of Message (A+B) were reported in April, including but not limited to English, Indonesian, Filipino and Tibetan. The lifespan of this "drink salted water" message was rather long, as the another famous fact-checking platform in Taiwan, MyGoPen \footnote{https://mygopen.com/}, released an article to disprove this false medical advice again in October 2020 \footnote{https://mygopen.com/2020/10/salt-water.html}, 7 months after it was first seen in our dataset.

\begin{table}[H]
    \centering
\begin{tabular}{lp{0.5\linewidth} p{0.4\linewidth}}
\toprule
       &English Translation & Original\\
\midrule
(A) & 
This is a $100\%$ accurate information... Why did we see a huge decline of confirmed cases in China during the last few days? They simply forced their citizens to rinse mouths with salted water $3$ times a day and then drink water for 5 minutes. The virus would attack throats before the lungs, and when getting in touch with salted water, the virus would die or get destroyed in lungs. This is the only way to prevent the spread of COVID-19. There is no need to buy medicine as there is nothing effective on the market.
& 
這是$100$\%準確的信息...
 為什麼中國過去幾天大大減少了感染人數？
 他們只是簡單地強迫他們的人民每天漱口3次鹽水。 完成後，喝水5分鐘。
 因為該病毒只能在喉嚨中侵襲，然後再侵襲肺部，當受到鹽水侵襲時，該病毒會死亡或從胃中流下來並在胃中銷毀，這是預防冠狀病毒流行的唯一方法。市場上沒有藥品，所以不要購買
\\
\midrule
(B) & Before reaching the lungs, the Novel Coronavirus would survive in throats for four days.  At this stage, people would experience sore throats and start coughing. If one can drink as much warm water with salt and vinegar, the virus could be destroyed. Share this information to save people's lives. &
新冠肺炎在還沒有來到肺部之前，它會在喉嚨部位存活4天。在這個時候，人們會開始咳嗽及喉痛。如果他能儘量喝多溫開水及鹽巴或醋，就能消滅病菌。儘快把此訊息轉達一下，因爲你會救他人一命！ \\
\midrule
(A+B) & Why did Mainland China show a huge decline of confirmed cases over the last few days?  Besides wearing masks and washing hands, they simply rinse mouths with salted water $3$ times a day and then drink water for 5 minutes [...] Dr. Wang of Tung Hospital stated that the Novel Coronavirus would survive in throats for four days before reaching the lungs [...] If one can drink as much warm water with salt and vinegar, the virus could be destroyed.[...]  &
為什麼中國大陸過去幾天大大減少了感染人數？除了戴口罩勤洗手外，他們只是簡單地每天漱口3次鹽水。 完成後，喝水5分鐘[...] 
新冠肺炎在還沒有來到肺部之前，它會在喉嚨部位存活4天[...] 如果他能儘量喝多溫開水及鹽巴或醋，就能消滅病菌[...] \\

\bottomrule
\end{tabular}
    \caption{Case 2 Message Content}
    \label{tab:c2-content}
\end{table}

\subsubsection{Case 3: This is a critical period, here are some suggestions...}

\begin{table}[H]
    \centering
\begin{tabular}{p{0.5\linewidth} p{0.4\linewidth}}
\toprule
       English Translation & Original\\
\midrule
10 days from now, Taiwan is in a critical period combating COVID-19. Here are some suggested measures.

1. Strictly prohibited going to public places.
2. Choose takeout from restaurants. 
3. Eat outside in open spaces. 
4. Wash your hands the right way (extremely important). 
5. When taking subway or bus, choose the seats at the first half of the vehicle. 
6. Do not wear contact lenses. 
7. Eat warm food and more vegetables. 
8. Avoid constipation. 
9. Drink warm water. 
10. Do not visit hair salons. 
11. Hang the clothes you're wearing outside for two hours the first thing you get home. 
12. Do not wear jewelry. 
13. Wash your hands immediately after touching cash or coins. Put coins you just received inside a plastic bag for one day before using them. 
14. Do not use colleague's phone when working. Disinfect before you have to use one. 
15. Avoid taking public transportation during rush hour. 
16. Do not visit night market or traditional market. 
17. Exercise. 
18. Avoid going to the gym.
& 今天開始10天，
台灣正式進入
武漢肺炎関鍵期。
建議如下:
1.嚴禁進入公共場所。
2.用餐儘量將食物外帶。
3.用餐環境儘量在 戶外。
4.正確方式的洗手(特別重要)。
5.坐捷運(公車)，選擇在車前頭。
6.避免戴隱形眼鏡
7.吃熱食,避開生凉食物,多吃蔬菜
8.保持腸胃顺暢。
9.多喝溫水。
10.暫停去髮廊。
11.穿過的衣服(外套,長褲),回家先單獨吊在戶外2小時
12.暫停戴首飾。
13.一有接觸錢幣,一定要洗手,剛拿進來的錢弊,先單獨放在塑膠袋中,一天後,才拿出來.
14.在公司不要使用別人的電話筒。電話筒的消毒。
15.避開巔峰時間坐車.
16.不去傳統市場及夜市.
17.適當的運動。
18.暫停進入健身房。
\\
\bottomrule
\end{tabular}
    \caption{Case 3 Message Content}
    \label{tab:c3-content}
\end{table}

This rumors first appeared in the dataset on February $6^{th}$ and has a total of 2121 reports in our dataset. Over the 1.5 months of its most popular time, it went viral at least two times: one on February $17^{th}$ with 394 reports, and on March $19^{th}$ with 543 reports. It was fact-checked by the Taiwan FactCheck Center on February $15^{th}$, 2020 \footnote{https://tfc-taiwan.org.tw/articles/2547}, however, the fact-check did not avoid the message from getting attention. The content started with authoritative tone that announced "We are at the most critical period of COVID-19", and then provided a list of "do's and dont's". While some \textit{suggestions} made medical sense in terms of hygiene, others didn't \footnote{https://tfc-taiwan.org.tw/articles/2547}. 
It was not stated explicitly in the message what the critical period was referring to, however, when taking together the listed "guidelines" into account, we could deduce that it hinted at the "critical period to prevent community spread". Community spread (社區感染）is a phase in a pandemic where many people who tested positive in an area cannot be determined how they got infected \footnote{https://www.cdc.gov/coronavirus/2019-ncov/faq.html\#Spread}. It is not hard to imagine that people would be concerned and worried about this significant phase where the risk of getting infected is greatly increased. In fact, we observed that such concerns co-occurred with the spread of this piece of message in February. 

\begin{figure}[H]
    \centering
    \includegraphics[width = 1\textwidth]{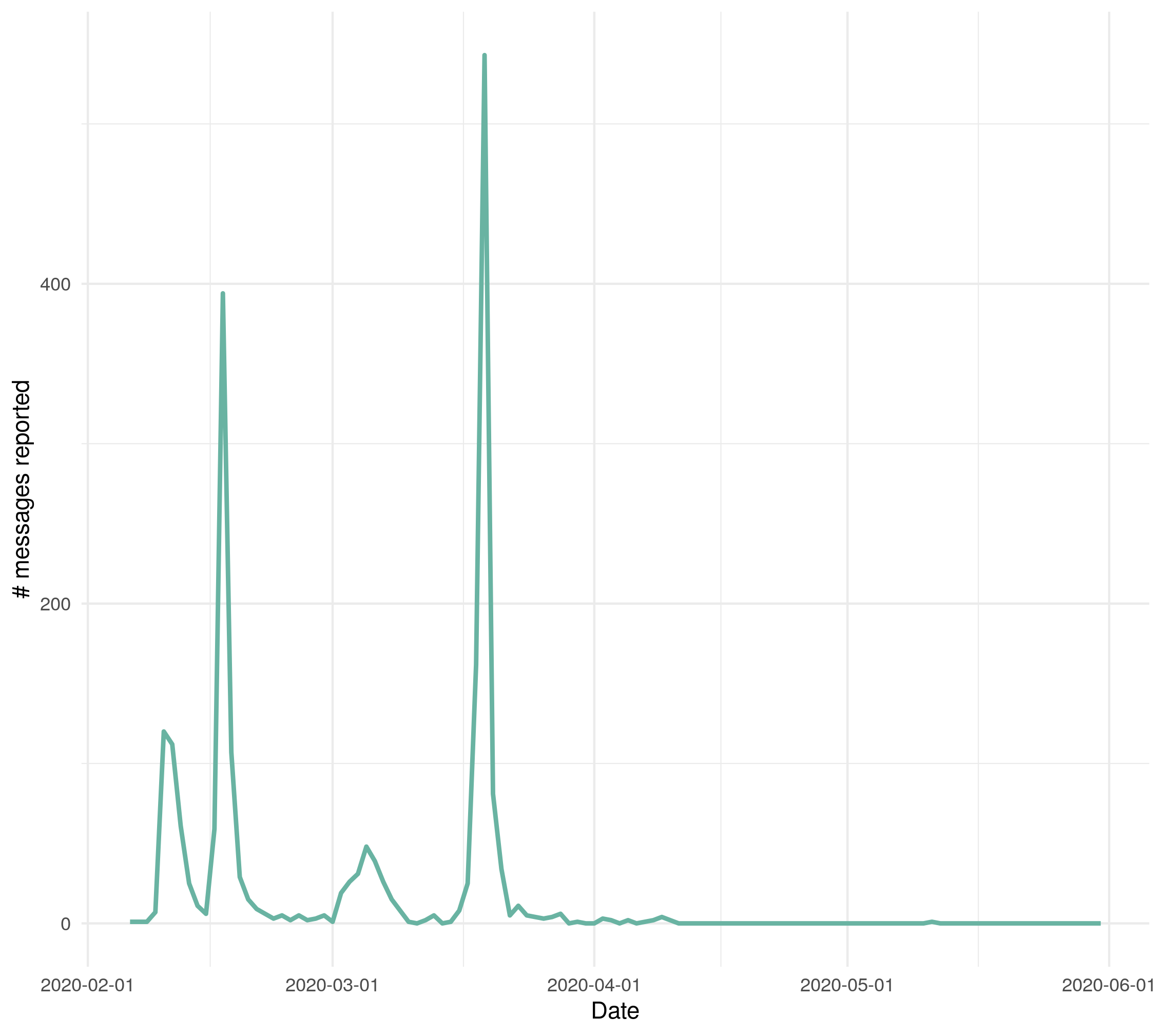}
    \caption{number of documents of Case 3 reported by date. The higher peaks were on Feb $17^{th}$ with 394 reports and March $19^{th}$ with 543 reports.}
    \label{fig:c3}
\end{figure}

\begin{figure}[H]
    \centering
    \includegraphics[width = 1\textwidth]{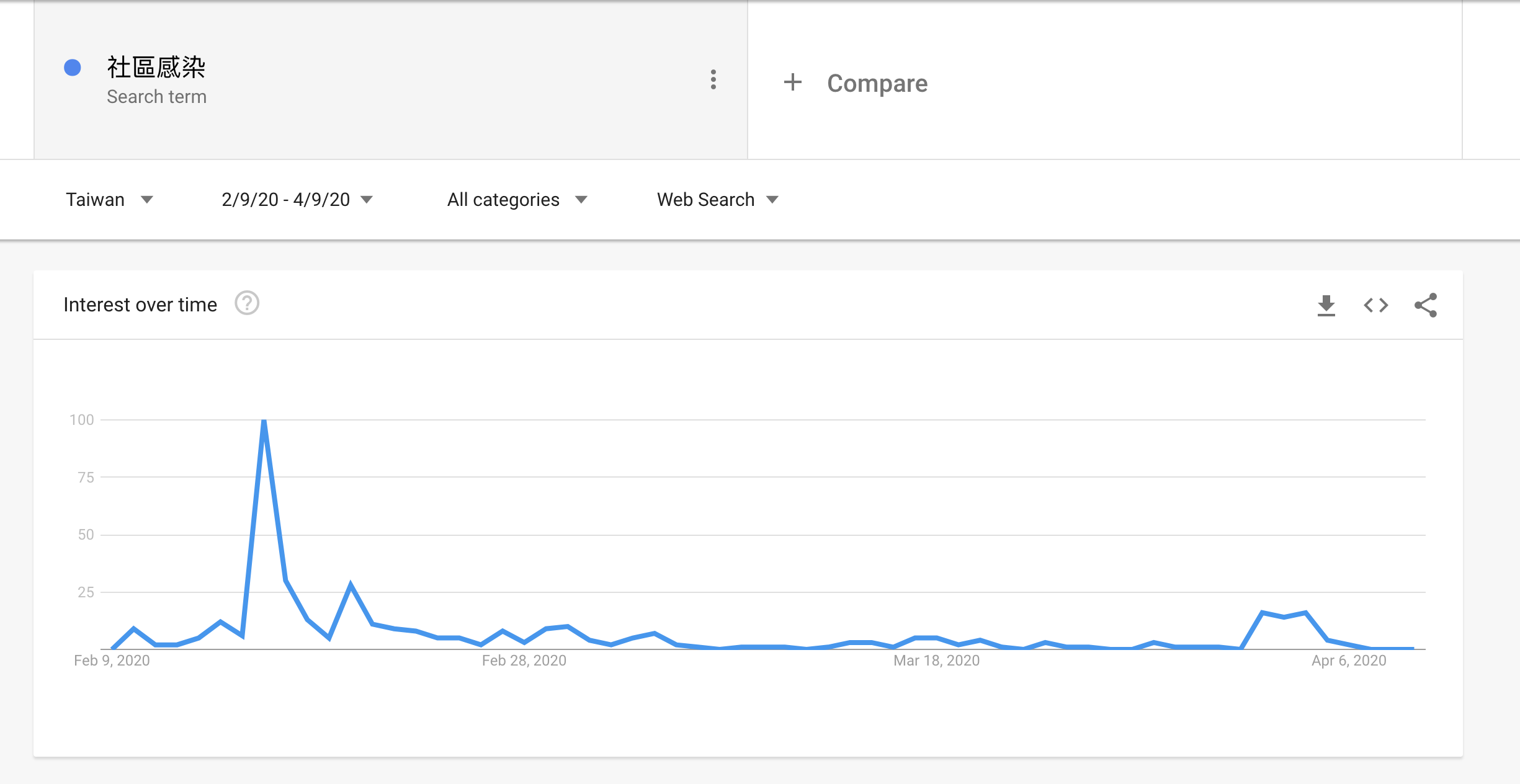}
    \caption{Google Trend of the interest in "community spread (社區感染)" in Taiwan between Feb $9^{th}$, 2020 and Apr $9^{th}$, 2020. The interest showed a sharp increase from Feb $15^{th}$ to Feb $16^{th}$, where it peaked.}
    \label{fig:community-spread-gtrend}
\end{figure}

On February $15^{th}$, 2020, Taiwan's Central Epidemic Command Center (CECC) reported that a taxi driver, infected by a person traveled back from China, was tested positive with the virus. He died on the same day and became the first death case in Taiwan. Over the next 4 days, four of his family members were also tested positive, forming the first COVID-19 cluster in Taiwan. During that time, people's concerns for community spread was looming.  In fact, Google trend for search term "社區感染 (Community Spread)" sharply increased on February $16^{th}$ (Figure \ref{fig:community-spread-gtrend}). Also, during this period, the number of the reported messages sharply increased (Figure \ref{fig:c3}). 

Content-wise, like what we observed in the first two cases, authorities, especially medical personnel, were used in several versions of the same message to "endorse" the content (Table \ref{tab:c3-changelog}). We spotted a major revision of the message on Feb $12^{th}$, 6 days after the first report, where the 18 bullets were pruned to 14, and strong words were modified to gentler tone. Last but not least, the message added a signature of \textit{"Regards from Medical Association"} on the last line. This became the most widespread version afterwards. Out of the $394$ documents reported on Feb $17^{th}$, $333$ documents were of this version.  Another key event in content transformation occurred on March $18^{th}$. On March $18^{th}$, Chen, Shih-chung, the CECC director, went to the Legislative Yuan (similar to Congress in the US) to answer interpellation about COVID-19. On the same day, messages started to have "Chen, Shih-Chung explained in the Legislative Yuan on March $18^{th}$ (3/18陳時中立法院說明)" before giving the list of \textit{suggestive measures}.  The next day, we saw another sharp increase of reported messages, reaching the highest peak. Of the $543$ messages reported on March $19^{th}$, $280$ has quoted Chen.

\begin{table}[H]
    \centering
\begin{tabular}{{lp{0.4\linewidth} p{0.4\linewidth}}}
\toprule
       Date &  Previous & New\\
\midrule
Feb. 12, 2020 & 1. Strictly prohibited going to public places.

1.嚴禁進入公共場所。& 1. Reduce going to public places. 

1.減少進入公共場所。 \\
\cmidrule{2-3}
&
3. Eat outside in open spaces.

5. When taking subway or bus, choose the seats at the first half of the vehicle. 

10. Do not visit hair salons.

16. Do not visit night market or traditional market.

3.用餐環境儘量在 戶外。

5.坐捷運(公車)，選擇在車前頭。

10.暫停去髮廊。

16.不去傳統市場及夜市.
& 
\textit{deleted}\\
\cmidrule{2-3}
&&
Regards from Medical Association

醫師全聯會關心 您\\

\midrule

Mar. 18, 2020 & 
10 days from now, Taiwan is in a critical period combating COVID-19. Here are some suggested measures.

今天起10天，台灣正式進入武漢肺炎関鍵期，
建議如下 & 

10 days from now, Taiwan is in a critical period combating COVID-19 \textbf{(Explained by Chen, Shi-Chung in Legislative Yuan on March $18^{th}$)}. Here are some suggested measures.

今天起10天，台灣正式進入武漢肺炎関鍵期，(3/18陳時中立法院說明)
建議如下\\

\bottomrule
\end{tabular}
    \caption{Content Change Log for Case 3}
    \label{tab:c3-changelog}
\end{table}

\section{Discussion}  \label{sec: discussion}

Similar to the findings of \cite{wood2019elusive}, we found that fact-check did not effectively alleviate the spread of false information. The popularity of rumors were more associated with major societal events or content changes. In addition to the above 3 case studies, we went through five other COVID-19 related rumors and manually identified common patterns of textual changes in their propagation. First of all, we observed that key authoritative figures were often (falsely) mentioned or quoted. For example, COVID-19 rumors often included medical-related persons, such as doctors or head of CECC. In addition, it was quite common to observe messages having a line or two disclaimers that expressed the uncertainty of truthfulness of the forwarded messages. For example, \textit{The following is for your reference only, I do not guarantee the truthfulness of the message. (以下謹提供參考不代表是否正確)} was seen in some messages during propagation. Many messages also included simplified Chinese characters or terms that are rarely used in Taiwan. For example, while in Taiwan, people refer to SARS pandemic as "SARS", a large number of messages use "非典", which is a term more popularly used in China. We also noticed messages that were a merge of other previously independent ones, and messages that included translation to other non-Chinese languages.  

These characteristics could serve as rules to discover possible false information as early detection mechanism. Although we identified these characteristics manually this time, it is quite possible to employ techniques such as Natural Language Processing to automatically recognize these textual changes in the future, making it possible to have a automatic early warning system of misinformation that does not involve fact-check by professionals. 

This study had several limitations. First, this data was collected by people's reports. Therefore, it was impossible to infer the true distribution of messages without making some assumptions. That is, if we saw more health-related misinformation in our data, it did not necessarily translate to more health-related rumors circulating in the platform.  In fact, it could also be that people were more alerted and skeptical at truthfulness health-related information. In addition, we only looked at text messages, therefore, information distributed visually or in audio was not covered. Lastly, our algorithm to group messages does not work well with short texts.

\section{Conclusion}  \label{sec: conclusion}
In this paper, we analyzed COVID-19 related rumors on a closed-messaging platform, LINE. We proposed a clustering algorithm that reduced the computational time from exponential to linear time. The algorithm enabled us to investigate the evolution of text messages. In fact, the algorithm enabled the research community to perform large-scale studies on the evolution of text messages at message-level rather than topic-level.
Similar to what \cite{shin2018diffusion} discovered in its study of $17$ political rumors, we found that false COVID-19 rumors tend to resurface multiple times even after being fact-checked, and with different degrees of content alterations. Furthermore, the messages often falsely quoted or mentioned authoritative figures, and such practice was helpful for the rumor to reach broader audiences.  Also, the resurfacing patterns seemed to be influenced by major societal events and content change. However, each peak of popularity would not last long and it was often without good explanation about how one wave of propagation ended. 
To the best of our knowledge, this is one of the few works that study COVID-19 misinformation on closed-messaging platforms and the first to study textual evolution of COVID-19 related rumors during its propagation. We would hope that this would further spark more studies in rumor propagation patterns.
\printbibliography
\end{document}